\newif\ifproofs

\newif\ifFullVersion

\documentclass[9pt,final,twocolumn]{IEEEtran}
\IEEEoverridecommandlockouts
\usepackage{balance}
\usepackage{amsmath}
\usepackage{amsthm}
\usepackage[bookmarks,colorlinks]{hyperref}
\usepackage{cite}
\usepackage{cases}
\usepackage{amssymb}
\usepackage{algorithm}
\usepackage{algorithmic}
\usepackage{graphicx}
\usepackage{subcaption}
\usepackage{color}
\usepackage{enumerate}
\usepackage{xcolor}
\usepackage{acronym}

\acrodef{adc}[ADC]{analog-to-digital convertor}
\acrodef{bilispa}[BiLiSPA]{bit-limited sparse linear array}
\acrodef{dft}[DFT]{discrete Fourier transform}
\acrodef{cs}[CS]{compressed sensing}
\acrodef{sla}[SLA]{sparse linear array}
\acrodef{ula}[ULA]{uniform linear array}
\acrodef{doa}[DOA]{direction of arrival}
\acrodef{had}[HAD]{hybrid analog and digital}
\acrodef{mse}[MSE]{mean square error}
\acrodef{sdp}[SDP]{semi-definite programming}
\acrodef{mimo}[MIMO]{multiple input and multiple output}
\acrodef{evd}[EVD]{eigenvalue decomposition}
\acrodef{rnn}[RNN]{recurrent neural network}
\acrodef{cnn}[CNN]{convolutional neural network}
\acrodef{nn}[NN]{neural network}

\newcommand{\figWidth}{2.3in}
\newcommand{\figHeight}{1.25in}

\DeclareMathOperator{\diag}{diag}

\DeclareMathOperator{\sign}{sign}

\title{\huge TransMUSIC: A Transformer-Aided Subspace Method for DOA Estimation with Low-Resolution ADCs}
  
\author{Junkai~Ji, Wei~Mao, Feng~Xi, Shengyao~Chen
\thanks{J. Ji, W. Mao, F. Xi, and S. Chen are with the Department
of Electronic Engineering, Nanjing University of Science and Technology, Nanjing 210094, China (email: xifeng@njust.edu.cn). 
This work was partly supported by the Natural Science Foundation of Jiangsu Province, China, under grant No.BK20221486. }
\vspace{-1.0cm}
}
\begin{document}
%
\maketitle
\pagestyle{empty}  
\thispagestyle{empty} 
\begin{abstract}
 Direction of arrival (DOA) estimation employing low-resolution \acp{adc} has emerged as a challenging and intriguing problem, particularly with the rise in popularity of large-scale arrays.
 The substantial quantization distortion complicates the extraction of signal and noise subspaces from the quantized data.
 To address this issue, this paper introduces a novel approach that leverages the Transformer model to aid the subspace estimation.
 In this model, multiple snapshots are processed in parallel, enabling the capture of global correlations that span them.
 The learned subspace empowers us to construct the MUSIC spectrum and perform gridless DOA estimation using a neural network-based peak finder.
 Additionally, the acquired subspace encodes the vital information of model order, allowing us to determine the exact number of sources.
 These integrated components form a unified algorithmic framework referred to as TransMUSIC. 
 Numerical results demonstrate the superiority of the TransMUSIC algorithm, even when dealing with one-bit quantized data. 
 The results highlight the potential of Transformer-based techniques in DOA estimation.
\end{abstract}
\vspace{-0.3cm}
\section{Introduction}
\label{sec:intro}
\vspace{-0.1cm}
The task of \ac{doa} estimation plays an essential role in a wide range of applications, including radar and sonar positioning, wireless communication, sensor networks, and so on \cite{Krim-SPM96,Trees2002OptimumAP}. 
A variety of algorithms have emerged to deduce DOAs from single or multiple snapshots of received signals.
These encompass classical subspace-based techniques \cite{Schmidt-86, Stoic-MUSIC89,Rao-RMUSIC89,Roy-ESPRIT89}, as well as compressed sensing (CS)-based methods \cite{Malioutov-TSP05, Yin-TSP11, Stoica-SPICE11, Yang-TSP14}.
Notably, while each snapshot denotes a discrete-time sample of the received signal, conventional DOA estimation methods often overlook the influence of quantization, which discretizes signal amplitude. 
Most current array systems resort to high-resolution \ac{adc}s to minimize the induced quantization distortion.
However, this practice comes at the cost of increased expenses, power consumption, and substantial data volume, particularly concerning large-scale arrays.
Therefore, it poses a significant challenge to develop an efficient DOA estimation method using low-resolution \ac{adc}s, which explicitly takes account of the effect of quantization distortion.

In the context of \ac{doa} estimation using low-resolution 
\ac{adc}s, the significant quantization distortion complicates the feasibility of employing subspace-based methods. 
While techniques like the arcsin law \cite{VanVleck-66}, also known as the Bussgang theorem \cite{Bussgang-52}, can be used to reconstruct the unquantized normalized covariance matrix from one-bit quantized data, thereby facilitating the use of subspace-based methods as demonstrated in \cite{Bar-Shalom-TAES02,Liu-ICASSP17,Huang-SPL19,Sedighi-ICASSP20}, a large performance gap exists between these one-bit \ac{doa} estimation methods and their classic counterparts.
Furthermore, using the time-varying sampling thresholds will significantly complicate the deriving of the covariance matrix from one-bit quantized data \cite{Eamaz-TSP22}.
Moreover, it is challenging to extend the Bussgang theorem to cases involving arbitrary low-bit quantizers.
Another approach to perform \ac{doa} estimation with low-resolution \ac{adc}s relies on quantized CS theory \cite{Zymnis-SPL10,Laska-TSP11}, which exploits the inherent sparsity of the signals and recovers the sparse solutions through convex optimization \cite{Yu-SPL16}.
While the covariance fitting techniques \cite{Xi-TSP20,Xi-TAES20,Xia-CIE21,Sedighi-SPL21} partially address the intrinsic off-grid issue associated with CS-based methods, the surge in computational complexity imposes limitations on their practical applications.

The recent development of deep learning has paved the way for the emergence of diverse data-driven \ac{doa} estimation methods \cite{Elbir-SL20,Liu-TAP18,Papageorgiou-TSP21,Wu-SPL19}.
These approaches leverage multi-layer \acp{nn} to learn the intricate relationship between \ac{doa}s and the received data or the covariance matrix by training on a given data set.
While such black-box neural networks enable the approximation of complex nonlinear mappings, they do so without utilizing the underlying knowledge of the \ac{doa} model, lacking the interpretability of those classic model-based methods.
These models are typically trained on massive datasets, limiting their adaptability when confronted with different scenarios.

Building upon the paradigm of model-based deep learning \cite{Shlezinger-DSLW21,Shlezinger-23}, recent works \cite{Merkofer-ICASSP22,Shmuel-ICASSP23,merkofer-damusic,shmuel-subspacenet} have introduced a hybrid approach that combines both model-based and data-driven approaches for \ac{doa} estimation.
The strength of this hybrid approach lies in its ability to deal with the model mismatch problem by resorting to data-driven techniques, while simultaneously exploiting the model knowledge to simplify the \ac{nn}-based architecture.
For example, the DA-MUSIC algorithm \cite{Merkofer-ICASSP22,merkofer-damusic} first employs a \ac{rnn} to learn a surrogate covariance matrix from the received data,
then seamlessly integrates this learned information with the established structure of the classic MUSIC algorithm to estimate the \ac{doa}s.
This combination empowers the data-driven aspect to successfully overcome the model mismatch issues that might arise when the actual signal model deviates from the assumed one.
It is worth noting that while these hybrid approaches present a novel pathway for developing \ac{doa} estimation methods, they mainly focus on constructing data-driven models to learn so-called covariance matrices, which are then used to estimate subspaces via \ac{evd}.
Thus, subspace estimation still heavily relies on model-based methods, which is a computationally intensive task.

This paper is dedicated to developing a hybrid approach to recover high-resolution \ac{doa}s from the data involving significant quantization distortion.
Specifically, we use a Transformer module \cite{NIPS2017} to directly learn the noise subspace from the quantized data by exploiting the attention mechanism to capture underlying correlations present in the data.
In contrast to the \ac{cnn} or \ac{rnn}-based architectures, the Transformer model processes multiple snapshots in parallel, enabling to capture long-range information among them.
We combine the Transformer-based subspace estimator with the algorithmic streamline of the classic MUSIC algorithm and yield a comprehensive framework, referred to as TransMUSIC.
By leveraging the estimated subspace, we can construct the MUSIC spectrum, facilitating the localization of \ac{doa}s by finding its peaks.
Furthermore, the learned subspace, in essence, encapsulates valuable information regarding the number of sources, which in turn enables us to deduce the source numbers from it.
Therefore, TransMUSIC establishes a unified framework capable of not only recovering \ac{doa}s but also estimating the model order, thereby eliminating the need for prior knowledge of the exact number of sources.
Numerical results show that even with just one-bit quantized data, TransMUSIC can outperform the classic MUSIC algorithm utilizing unquantized data. This remarkable performance proves the effectiveness of the attention mechanism in extracting the critical features for accurate \ac{doa} estimation.


\vspace{-0.2cm}
\section{Signal Model and Motivations}
\label{sec:model}
\vspace{-0.1cm}
In this section, we first describe the array signal model and the quantization operation, then briefly introduce the classic MUSIC algorithm, and finally discuss the motivation of this work.

\vspace{-0.2cm}
\subsection{Signal Model}
\label{subsec:signal_model}
\vspace{-0.1cm}
Consider a scenario with $K$ narrowband far-field source signals impinging on a linear array with $M$ omnidirectional sensors from different directions $\theta_k\in [-90^{\circ},90^{\circ}]$. The locations of receive antennas are given by $\{r_{1}d,r_{2}d,\cdots,r_{M}d\}$, where $r_{1}<r_{2}<\cdots<r_{M},r_{m}\in\mathbb{Z}$ and $d=\frac{\lambda}{2}$ with $\lambda$ denoting the wavelength of sources. 
For simplicity, we assume a \ac{ula} used in this paper, i.e., $r_m = m-1$.
Then the $L$ snapshots received at the array can be represented by
\begin{equation}
  \begin{split}
    \mathbf{y}(t)&=\sum_{k=1}^{K} {\mathbf{a}(\theta_k)x_k(t)}+\mathbf{n}(t)\\
    &=\mathbf{A}(\boldsymbol{\theta}) \mathbf{x}(t)+\mathbf{n}(t), \quad t=1,2,\cdots,L
  \end{split}
\label{eqn:yt}
\end{equation}
where $\mathbf{y}(t),\mathbf{n}(t)\in \mathbb{C}^M$, and $\mathbf{x}(t)\in \mathbb{C}^K$ denote the received signals, additive noises, and source signals, respectively. 
$\mathbf{A}(\boldsymbol{\theta})=[\mathbf{a}(\theta_1),\mathbf{a}(\theta_2),\cdots,\mathbf{a}(\theta_K)]\in \mathbb{C}^{M\times K}$ is the steering matrix with $\boldsymbol{\theta}=\{\theta_1, \theta_2, \cdots, \theta_k\}$ and $\mathbf{a}(\theta_k)=[e^{j2\pi\frac{r_{1}d\sin\theta_k}{\lambda}},\cdots,e^{j2\pi\frac{r_{M}d\sin \theta_k}{\lambda}}]$.

Before performing digital processing, the received signal $\mathbf{y}(t)$ has to be converted to a set of digital representations by using \ac{adc}s.
Classic \ac{doa} estimation methods typically assume the use of high-resolution \ac{adc}s, effectively disregarding the influence of quantization distortion.
However, in this paper, we explicitly consider the effect of quantization distortion and apply only 1-bit quantization to the observed signal.
Specifically, we use a pair of one-bit ADCs to independently quantize the real and imaginary components of the received signal.
Let $\mathcal{Q}_1(\cdot) = \frac{1}{\sqrt{2}}(\sign(\Re\{\cdot\}) + j\sign(\Im\{\cdot\}))$ be the complex one-bit quantization operator, where $\sign(\cdot)$ denote the sign function applied element-wise to any vector or matrix.
Thus, the one-bit quantized data becomes
\begin{equation}
    \mathbf{z}(t) = \mathcal{Q}_1(\mathbf{y}(t)) = \mathcal{Q}_1(\mathbf{A}(\boldsymbol{\theta}) \mathbf{x}(t)+\mathbf{n}(t)).
\end{equation}

Therefore, the DOA estimation is to recover the set of angles $\boldsymbol{\theta}$ from the quantized data $\{\mathbf{z}(t)\}_{t=1}^L$.
Due to the low-resolution \ac{adc}s, we have to explicitly take the distortion induced by quantization into account and design the specific algorithm for DOA estimation.

\vspace{-0.2cm}
\subsection{SubSpace-based DOA Estimation}
\label{subsec:subspace}
\vspace{-0.1cm}
Subspace methods are a class of DOA estimation algorithms that hinge on precisely estimating both signal and noise subspaces derived from the received signals.
A representative algorithm within this category is the MUSIC algorithm \cite{Schmidt-86}, which deduces the subspaces and recovers the \ac{doa}s from the covariance matrix $\mathbf{R}_y\triangleq\mathbb{E}\{\mathbf{y}(t)\mathbf{y}^H(t)\}$.
The inferences in the MUSIC algorithm are based on the following assumptions:
\begin{itemize}
	\item[A1] The number of sources is known, and the source signals are uncorrelated, i.e., $\mathbf{x}(t)\sim\mathcal{CN}(\boldsymbol{0}, \mathbf{P})$, where $\mathbf{P}=\diag(p_1,p_2,\cdots,p_{K})$ denotes the covariance matrix of $\mathbf{x}(t)$;
	\item[A2] The additive noise are independently and identically distributed as $\mathcal{CN}(\boldsymbol {0}, \sigma^2\mathbf{I}_M)$, where $\sigma^2$ is the variance of noise;
 	\item[A3] There is no temporal correlation between each snapshot, and the additive noise is uncorrelated with the sources.
\end{itemize}
Then the noise and signal subspaces can be identified via eigenvalue decomposition (EVD) of the covariance matrix $\mathbf{R}_y=\mathbf{A}(\boldsymbol{\theta}) \mathbf{P}\mathbf{A}^H(\boldsymbol{\theta})+\sigma^2\mathbf{I}_M$.
By restricting the number of sources to be less than the number of antennas, i.e., $K<M$, the eigenvectors corresponding to the largest $K$ eigenvalues consist of the signal subspace, while the eigenvectors corresponding to the remaining $M-K$ eigenvalues consists of the noise subspace.
We can construct the following MUSIC spectrum:
\begin{equation}\label{eqn:P_MUSIC}
    P_{\text{MUSIC}}(\psi) = \frac{1}{\mathbf{a}^H(\psi)\mathbf{E}_n\mathbf{E}^H_n\mathbf{a}(\psi)},
\end{equation}
where $\mathbf{E}_n\in\mathbb{C}^{M\times (M-K)}$ denotes the noise subspace.
By exploiting the orthogonality of signal and noise subspaces, the DOAs are estimated by finding the $K$ peaks of $P_{\text{MUSIC}}(\psi)$.
In practical scenarios, finite-length snapshots necessitate the use of the sample covariance matrix (SCM) $\hat{\mathbf{R}}y=\frac{1}{L}\sum_{t=1}^L\mathbf{y}(t)\mathbf{y}^H(t)$ as a substitute for $\mathbf{R}_y$. 

Nonetheless, when using the low-resolution \ac{adc}s, the data available is the quantized observations $\{\mathbf{z}(t)\}_{t=1}^L$ rather than the received signal $\{\mathbf{y}(t)\}_{t=1}^L$.
While it is possible to establish the relation between $\mathbf{R}_y$ and $\mathbf{R}_z \triangleq \mathbb\{\mathbf{z}(t)\mathbf{z}^H(t)\}$ using the arcsine law \cite{VanVleck-66} or the Bussgang theorem \cite{Bussgang-52}, the performance of subspace estimation is significantly degraded due to the intricate interplay between the quantized measurements and the desired signal and noise subspaces.
Consequently, the task of accurately deducing subspaces from quantized observations becomes a notable challenge, owing to its pivotal role in achieving precise high-resolution \ac{doa} estimation.

\vspace{-0.2cm}
\subsection{Motivations}
\label{subsec:Moti}
\vspace{-0.1cm}
A vital limitation of the classic MUSIC algorithm lies in its incapability to deduce the subspaces when the assumptions A1$\sim$A3 are violated. 
In cases where the number of sources is unknown, additional processing is required to determine the exact number of sources present in the signals \cite{Wu-ICASSP94,Qiu-ICCP15}.
Furthermore, quantization distortion impairs the covariance matrix's ability to capture the information of signal and noise subspaces precisely.
In practical scenarios, the SCM deviates significantly from the actual covariance matrix due to limited snapshots, further increasing inaccuracies in EVD-based subspace estimation.
These challenges motivate us to develop a data-driven method to establish a robust nonlinear mapping between the quantized data and the desired signal or noise subspace.

\vspace{-0.2cm}
\section{TransMUSIC}
\label{sec:method}
\begin{figure*}
\centering
\includegraphics[width=0.8\linewidth]{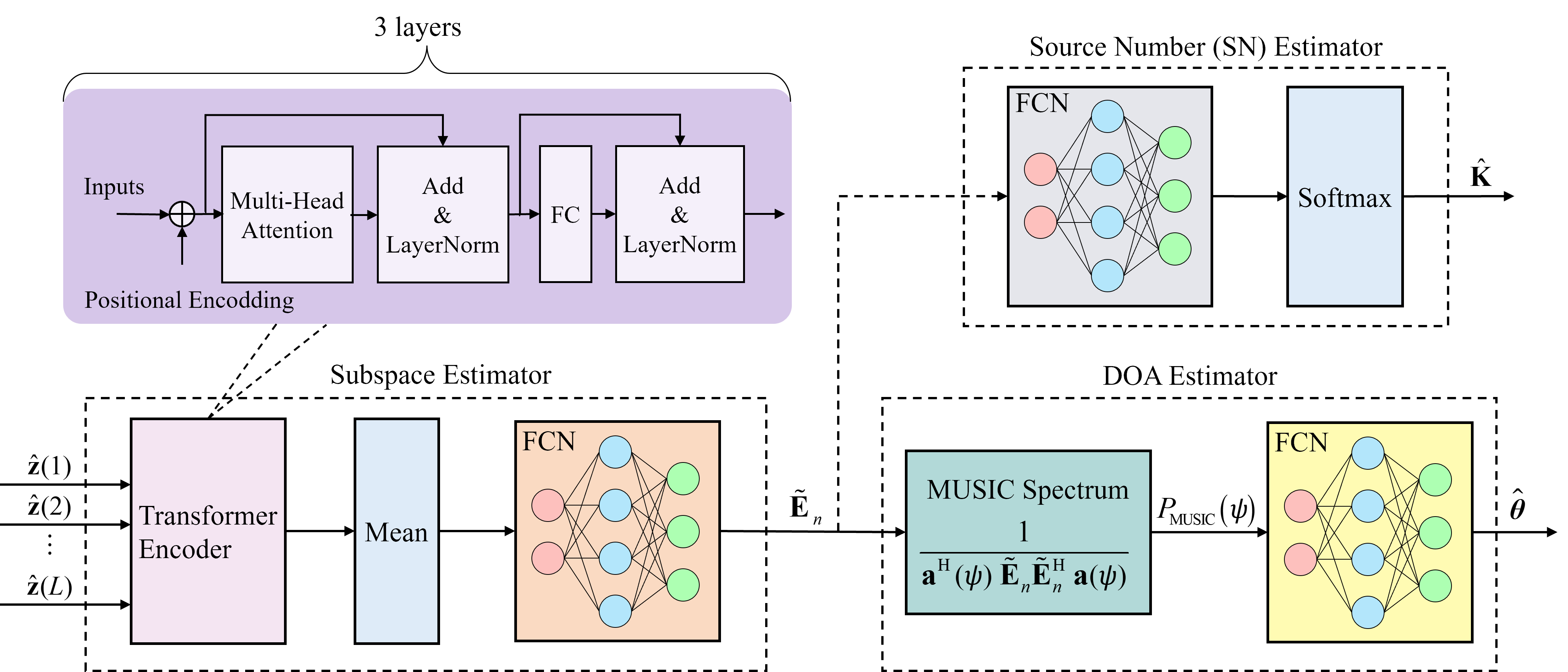}
\caption{The architecture of the TransMUSIC algorithm.}
\label{transmusic}
\end{figure*}

\vspace{-0.1cm}
In this section, we present the architecture of the proposed TransMUSIC, which builds upon the algorithmic framework of the classic MUSIC algorithm and replaces the EVD-based subspace estimation with a Transformer encoder.
The overall architecture of the TransMUSIC algorithm is shown in Fig. \ref{transmusic}, consisting of a subspace estimator, a \ac{doa} estimator, and a source number (SN) estimator.
In particular, the transformer encoder processes the $L$ snapshots in parallel to extract essential features of the quantized data.
These features contribute to subspace estimation through a dedicated fully connected network (FCN).
Then we can construct the MUSIC spectrum (\ref{eqn:P_MUSIC}) and estimate the \ac{doa}s facilitated by a peak-finding network based on another FCN.
A significant advantage of the TransMUSIC is its capability to perform subspace estimation without prior knowledge of the number of sources. Remarkably, the information about the source number is implicitly embedded within the estimated subspace, which can be extracted using an additional FCN branch.
We will elaborate on the architecture in the following subsection.

\vspace{-0.2cm}
\subsection{Architecture}
\label{subsec:archi}
\vspace{-0.1cm}
In the TransMUSIC, the transformer encoder employs the attention mechanism for effective feature extraction by concurrently processing the quantized observations $\{\mathbf{z}(t)\}_{t=1}^L$. 
To facilitate the application of the attention mechanism, we concatenate the real and imaginary parts of each snapshot $\mathbf{z}(t)$ into a vector $\hat{\mathbf{z}}(t) = [\Re(\mathbf{z}(t))^T, \Im(\mathbf{z}(t))^T]^T$, which then serves as the input to the transformer encoder.
Typically, the transformer encoder comprises a sequence of interconnected blocks, each of which encompasses a multi-head attention module, followed by a one- or two-layer FCN with an activation function (e.g., ReLU).
The output of the transformer encoder constitutes a collection of features, denoted as $\{\mathbf{o}(t)\}_{t=1}^L$, and these features maintain the same dimensionality as the input data.

Subsequently, these features are leveraged to formulate an estimate of the noise subspace, a crucial step in generating the MUSIC spectrum.
It is notable that in the ULA array with $M$ antennas, the covariance matrix $\mathbf{R}_y$ is a Hermitian and Toeplitz matrix.
This unique property allows for a parametric representation of the covariance matrix through an $M$-dimensional vector.
Based on this insight, an average unit is employed to compactly represent the extracted features into one feature vector $\bar{\mathbf{o}}=\frac{1}{L}\sum_{t=1}^L\mathbf{o}(t)$, facilitating subsequent processing independently of the input data's size.
Following this, an FCN network is employed to generate the desired noise subspace based on the feature vector.
Considering the dimension of the actual noise subspace $\mathbf{E}_n$ depends on the number of sources $K$ which is both varying and unknown, the FCN network outputs an augmented noise subspace $\tilde{\mathbf{E}}_n \in\mathbb{C}^{M\times M}$ in its place. 
Essentially, $\tilde{\mathbf{E}}_n$ can be viewed as a matrix with $M-K$ columns being the dominant ones.

After formulating the MUSIC spectrum (\ref{eqn:P_MUSIC}) using the augmented noise subspace $\tilde{\mathbf{E}}_n$, we use an additional FCN network to obtain the \ac{doa}s by finding the peaks of the spectrum.
The application of an FCN-based peak finder avoids the exhaustive search on the discrete grids, achieving superior resolution.
Addressing the dynamic output dimension of the FCN network, especially in diverse scenarios with varying numbers of sources, we resolve this by setting the output dimension to $M-1$. 
This choice aligns with the maximum number of sources a ULA with $M$ antennas can effectively resolve.
However, only the first $K$ outputs are considered as the estimated \ac{doa}s while the remaining outputs are ignored.
The derivation of $K$ will be addressed by the SN estimator.

The learned augmented noise subspace $\tilde{\mathbf{E}}_n$ inherently encapsulates the information about $K$, facilitating the determination of the number of sources.
For this purpose, we construct an SN estimator comprising an FCN network followed by a softmax layer, as illustrated in Fig. \ref{transmusic}.
Given that the maximum resolvable source number is $M-1$, the task is to resolve $M-1$ classifications. 
This naturally translates into a multi-class classification problem that can be effectively solved by using an FCN network.

\vspace{-0.2cm}
\subsection{Training Procedure}
\label{subsec:training}
\vspace{-0.1cm}
The TransMUSIC algorithm undergoes end-to-end supervised training.
The training set $\mathcal{D}$ consists of $D$ pairs of quantized observations and their corresponding \ac{doa}s, i.e., $\mathcal{D} = \left\{\left(\mathbf{Z}^{(d)},\boldsymbol{\theta}^{(d)}\right)\right\}_{d=1}^D$ with $\mathbf{Z}^{(d)}=[\mathbf{z}^{(d)}(1),\cdots,\mathbf{z}^{(d)}(L_d)]\in\mathbb{C}^{M\times L_d}$ and $|\boldsymbol{\theta}^{(d)}|=K_d$.
Let $\hat{\boldsymbol{\theta}}(\mathbf{Z}^{(d)})$ and $\hat{\mathbf{K}}(\tilde{\mathbf{E}}_N^{(d)})$ denote the $(M-1)$-dimension outputs of the \ac{doa} estimator and the SN estimator, respectively, applying to the data $\mathbf{Z}^{(d)}$ and the learned noise subspace $\tilde{\mathbf{E}}_N^{(d)}$.
To train the lower branch in Fig. \ref{transmusic} for \ac{doa} estimation, we employ the root mean squared periodic error (RMSPE) \cite{Routtenberg-TSP12,Routtenberg-TSP13} as the loss function:
\begin{equation}
    \mathcal{L}_{\text{RMSPE}}(\boldsymbol{\theta}^{(d)},\hat{\boldsymbol{\theta}}(\mathbf{Z}^{(d)})) = \text{RMSPE}(\boldsymbol{\theta}^{(d)},\hat{\boldsymbol{\theta}}_{1:K_d}(\mathbf{Z}^{(d)})),
\end{equation}
with $\text{RMSPE}(\boldsymbol{\theta},\hat{\boldsymbol{\theta}}) = \min_{\mathbf{P}\in\mathcal{P}_K}(\frac{1}{K}\|\text{mod}_{\pi}(\boldsymbol{\theta}-\mathbf{P}\hat{\boldsymbol{\theta}})\|^2)^{1/2}$, where $\mathcal{P}_K$ is the set of all $K\times K$ permutations and $\text{mod}_{\pi}$ denotes the element-wise modulus operation.
The SN estimator is trained using the cross-entropy as the loss function:
\begin{equation}
    \mathcal{L}_{\text{CE}}(K_d, \hat{\mathbf{K}}(\tilde{\mathbf{E}}_N^{(d)})) = -\log \hat{K}_{K_d}(\tilde{\mathbf{E}}_N^{(d)}),
\end{equation}
where $\hat{K}_{K_d}(\tilde{\mathbf{E}}_N^{(d)})$ denotes the $K_d$-th element of $\hat{\mathbf{K}}(\tilde{\mathbf{E}}_N^{(d)})$.
It is worth noting that the gradient propagation from the SN estimator to the subspace estimator should be blocked during backpropagation.

\vspace{-0.2cm}
\subsection{Discussion}
\label{subsec:discuss}
\vspace{-0.1cm}
Similar to the DA-MUSIC \cite{Merkofer-ICASSP22,merkofer-damusic}, the TransMUSIC we present here is built upon the algorithmic framework of the classic MUSIC algorithm. 
A key distinction between them lies in our utilization of a transformer encoder, which plays a central role in capturing the intricate nonlinear relationship between the quantized data and the noise subspace.
It performs this task through parallel processing, in contrast to the RNN in the DA-MUSIC algorithm, which sequentially analyzes the dependencies between consecutive snapshots.

Another notable advantage of TransMUSIC is that it directly estimates the noise subspace from the received data, circumventing the need for the computationally demanding EVD involved in subspace estimation from the covariance matrix.
This improvement significantly accelerates the overall \ac{doa} estimation procedure.
Moreover, due to its direct subspace estimation approach, TransMUSIC eliminates the necessity of using a selector to choose the corresponding eigenvectors, as done in the DA-MUSIC algorithm.
This simplifies the architecture and the training procedure.

While the motivation of our work is to develop an \ac{nn}-aided \ac{doa} estimation method addressing the effect of quantization distortion, it is worth noting that TransMUSIC's applicability is not confined to such scenarios.
The power of the transformer module facilitates the extension of our work to other non-ideal cases, such as coherent sources, arbitrary array geometry, and scenarios with some malfunctioning antennas.
We will leave these for future work.

\begin{figure}[htbp]
\centering
\includegraphics[width=\figWidth, height=\figHeight]{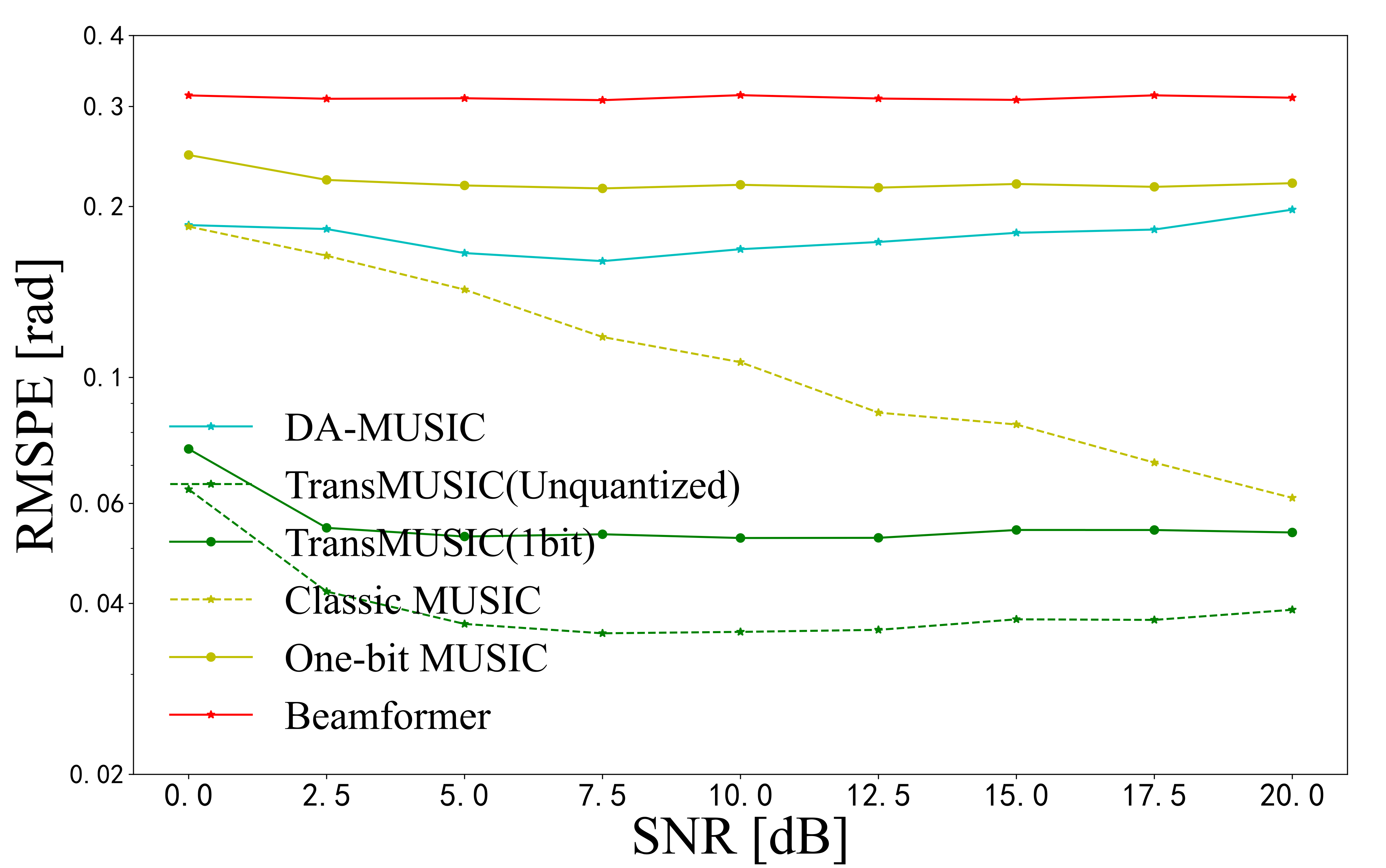} 
\vspace{-0.2cm}
\caption{RMSPE with respect to various SNRs.}
\label{RMSPEvsSNR}
\end{figure}

\begin{figure}[htbp]
\centering
\includegraphics[width=\figWidth, height=\figHeight]{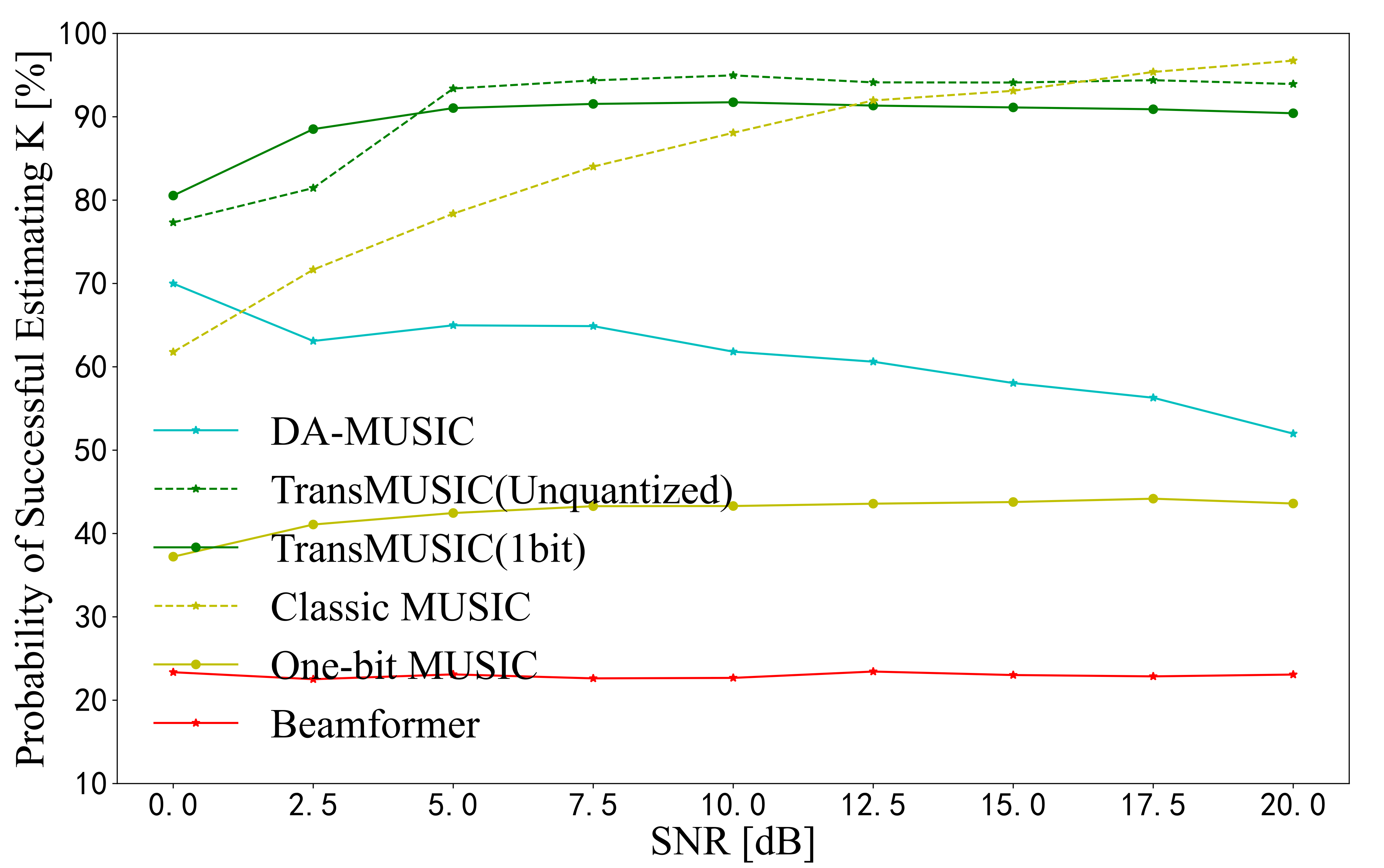} 
\vspace{-0.2cm}
\caption{Probability of successfully estimating $K$ versus various SNRs.}
\label{AccvsSNR}
\end{figure}

\begin{table}[!h]
\caption{Simulation Parameters}
\centering
\begin{tabular}{ |l|c|l|c| } 
 \hline
 Parameter & Value & Parameter & Value  \\ 
 \hline
  $M$ & 8& Training set size & 256000 \\ 
 $K$ & 2,3,4,5 & Validation set size & 25600 \\
 $L$ & 200 & Test set size & 25600\\
 SNR & 0dB,5dB,10dB & Batch size & 64 \\
 Optimizer & Adam & Initial learning rate & $10^{-3}$\\
 \hline
\end{tabular}
\label{Tab_I}
\end{table}

\vspace{-0.1cm}
\section{Numerical Experiments}
\label{sec:experiment}
\vspace{-0.1cm}
Now we evaluate the performance of the proposed TransMUSIC algorithm\footnote{The source code used in our experiment can be found online at \url{https://github.com/jijunkai/Transformer_Music}}.
Throughout the experiments, we consider a ULA with $M=8$ half-wavelength spaced elements.
The source signal $\mathbf{x}(t)$ and noise signal 
$\mathbf{n}(t)$ are both modeled as complex Gaussian distributions, in accordance with the assumptions A1$\sim$A3 in Subsection \ref{subsec:subspace}.
The \ac{doa}s $\boldsymbol{\theta}$ are randomly drawn from an i.i.d. uniform distribution over the range $(-\frac{\pi}{2},\frac{\pi}{2})$ and the number of sources is randomly distributed between $[2,5]$.
To facilitate our evaluations, we create synthetic datasets through simulations in different scenarios. 
Unless stated otherwise, the parameters used in our experiments are summarized in Table \ref{Tab_I}.

For the sake of comparison, we also implement several alternative approaches, including the classic MUSIC algorithm \cite{Schmidt-86}, the conventional beamformer method \cite{Bartlett-48}, and the DA-MUSIC algorithm \cite{merkofer-damusic}. Notably, all of these methods employ unquantized data as their input.
Furthermore, we also implement the one-bit MUSIC algorithm \cite{Huang-SPL19} utilizing one-bit quantized data.
Two TransMUSIC algorithms are implemented, training separately using one-bit quantized data and unquantized data.
With the unknown number of sources, the classic MUSIC algorithm and one-bit MUSIC algorithm determine the number of sources by estimating the multiplicity of the smallest eigenvalue while the beamformer method determines it by estimating the number of dominant peaks.


In the first experiment, we train the TransMUSIC and DA-MUSIC algorithms using the dataset mixed with SNRs of 0dB, 5dB, and 10dB, and then test the generalization of the algorithms across varying noise levels.
Fig. \ref{RMSPEvsSNR} illustrates the RMSPE of different algorithms with respect to various SNRs.
Both TransMUSIC algorithms, one with unquantized data and the other with one-bit data, consistently exhibit lower RMSPEs compared to the alternative methods.
Due to the diverse SNR scenarios used during training, the DA-MUSIC algorithm performs worse than the classic MUSIC algorithm, showing its limited adaptability to varying SNR conditions.
However, the TransMUSIC algorithm demonstrates robust performance even when applied to SNR scenarios that it was not explicitly trained on.
It is worth highlighting that TransMUSIC with one-bit data even outperforms those with unquantized data, underscoring the effectiveness of the Transformer architecture used for subspace estimation. 
We next evaluate the probability of successfully estimating the source number $K$ versus different SNRs, depicted in Fig. \ref{AccvsSNR}.
Again, both the TransMUSIC algorithms show better performance in identifying the exact number of sources across various SNR scenarios.
Conversely, the DA-MUSIC algorithm struggled to estimate it accurately. 
Its accuracy notably decreased when applied to SNR scenarios outside its training scope.

\begin{figure}[tbp]
\centering
\includegraphics[width=\figWidth, height=\figHeight]{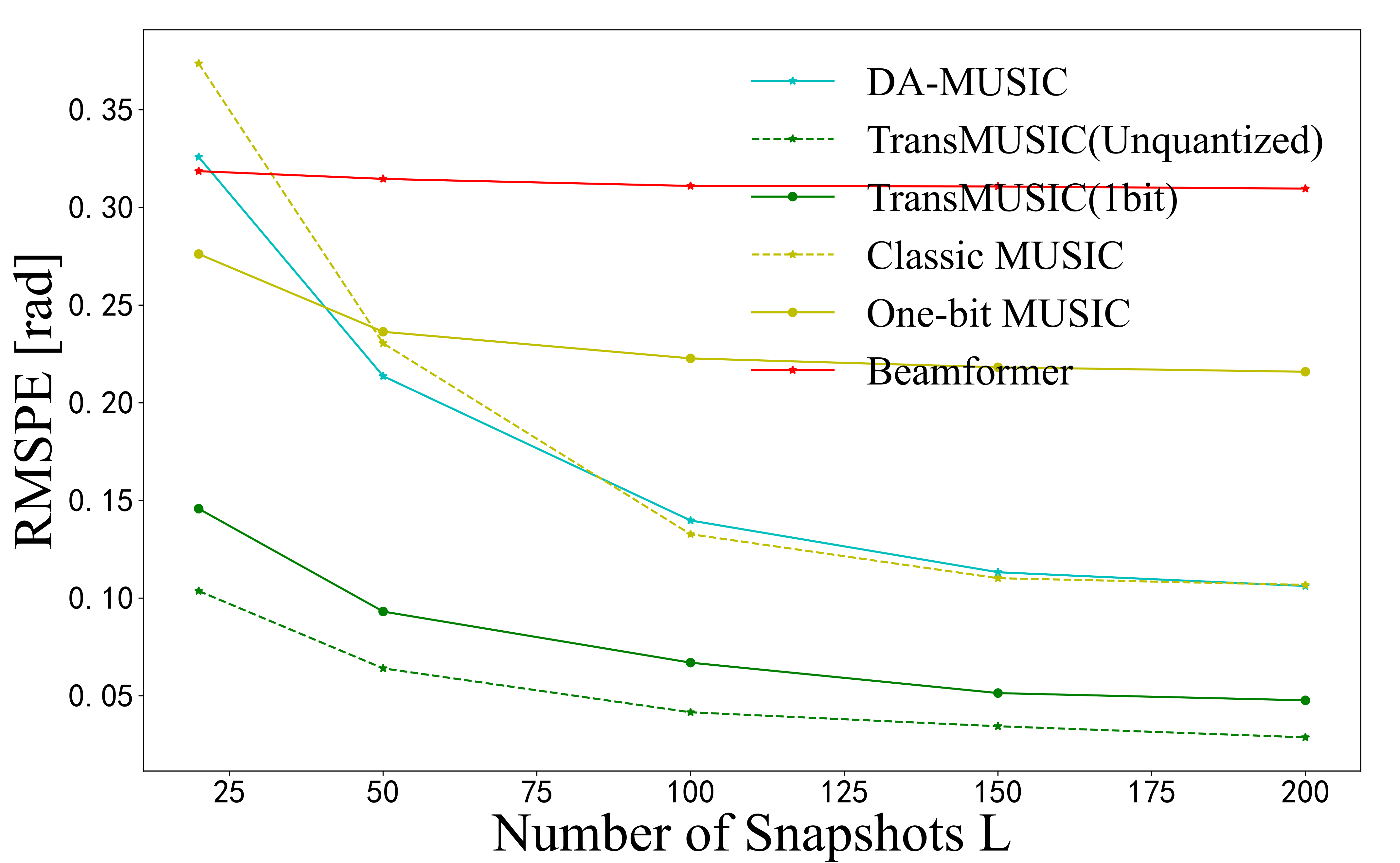} 
\vspace{-0.2cm}
\caption{RMSPE with respect to different number of snapshots $L$.}
\label{RMSPEvsL}
\end{figure}

\begin{figure}[tbp]
\centering
\includegraphics[width=\figWidth, height=\figHeight]{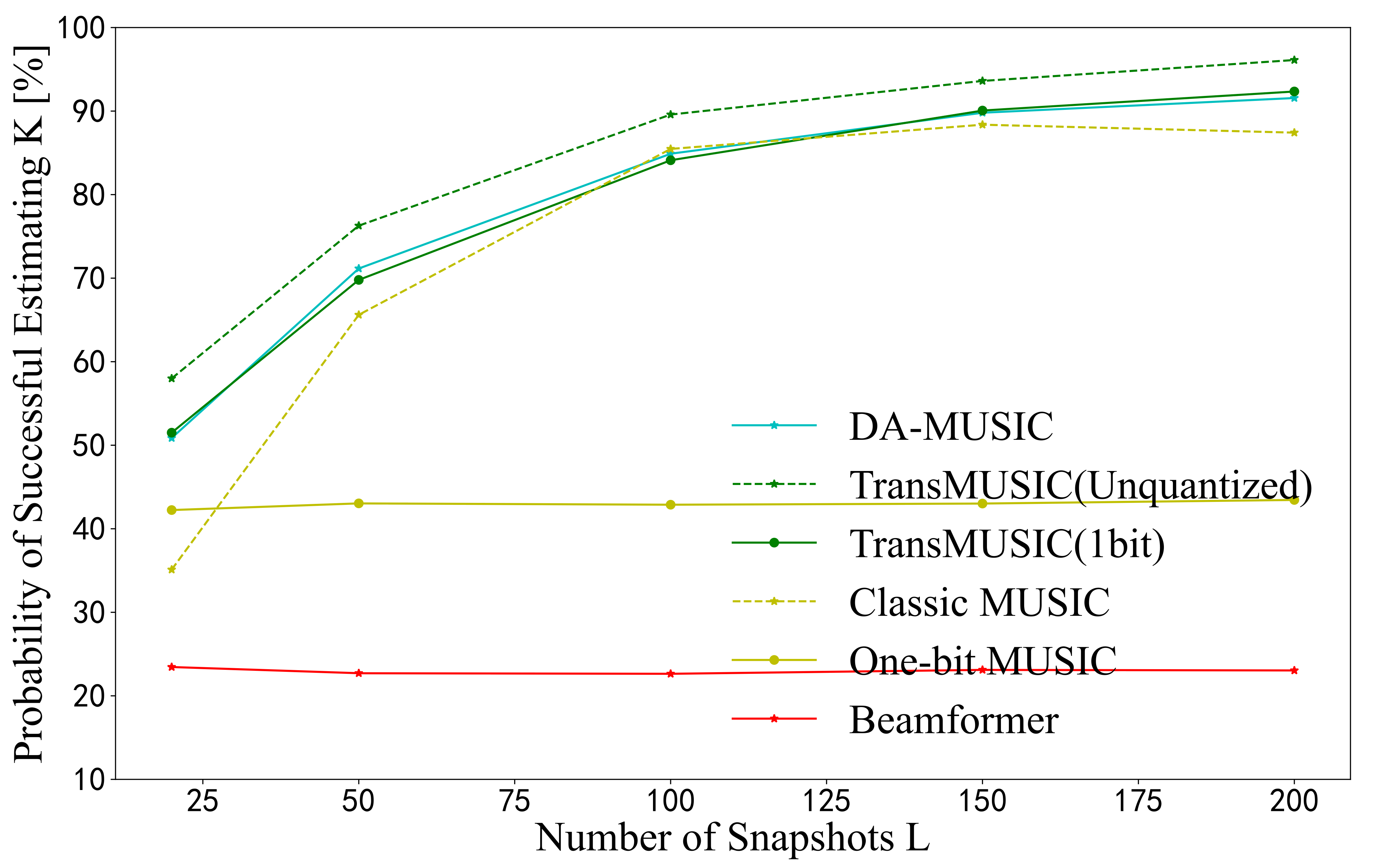} 
\vspace{-0.2cm}
\caption{Probability of successfully estimating $K$ versus different number of snapshots $L$.}
\label{AccvsL}
\end{figure}

In the next experiment, we maintain a fixed SNR of 10dB for training and testing.
Both TransMUSIC and DA-MUSIC algorithms are exclusively trained with $L=200$ snapshots and subsequently tested with varying $L$.
Fig. \ref{RMSPEvsL} presents the RMSPE of various algorithms with respect to different numbers of snapshots.
Although the performance degradation is observed with fewer snapshots, the TransMUSIC algorithms outperform their counterparts whether using unquantized data or using one-bit data.
It shows the superior performance of the TransMUSIC when implemented with only a few snapshots.
Unfortunately, as shown in Fig. \ref{AccvsL}, the probability of estimating $K$ significantly deteriorated with decreasing numbers of snapshots.
This result highlights the dependency of source number estimation on the number of snapshots available.
Addressing this shortcoming may require separate training for each number of snapshots.

\vspace{-0.2cm}
\section{Conclusion}
\label{sec:conclusion}
\vspace{-0.1cm}
In this work, we proposed a Transformer-aided subspace method, named TransMUSIC, for \ac{doa} estimation.
The powerful attention mechanism used in the Transformer module makes it possible to extract valuable features for subspace estimation, even in scenarios contaminated by significant quantization distortion.
TransMUSIC seamlessly integrates the Transformer-based subspace estimation into the algorithmic framework of the classic MUSIC algorithm. This not only simplifies the resulting architecture but also enhances the interpretability.
Simulation results demonstrate the superior performance of the TransMUSIC algorithm, regardless of the presence of quantization distortion.
Notably, the TransMUSIC algorithm, even when using one-bit data, surpasses the classic MUSIC algorithm and other counterparts that rely on unquantized data.
Experiments with real data will be our future work.

\bibliographystyle{IEEEtran}
\bibliography{IEEEabrv,reference}

\end{document}